# Low-Temperature Spectroscopic Investigation of Lead-Vacancy Centers in Diamond Fabricated by High-Pressure and High-Temperature Treatment

Peng Wang[1], Takashi Taniguchi[2], Yoshiyuki Miyamoto[3], Mutsuko Hatano[1], and Takayuki Iwasaki[1, *]

[1]Department of Electrical and Electronic Engineering, School of Engineering, Tokyo Institute of Technology, Tokyo 152-8552, JAPAN
[2]Research Center for Functional Materials, National Institute for Materials Science, Tsukuba 305-0047, JAPAN
[3]Research Center for Computational Design of Advanced Functional Materials, National Institute of Advanced Industrial Science and Technology, Tsukuba 305-8568, JAPAN

[*]e-mail: iwasaki.t.aj@m.titech.ac.jp

Key words:
Diamond, Lead-vacancy center, Single photon emitter, Polarization, HPHT annealing

Abstract
We report the optical observation of lead-vacancy (PbV) centers in diamond fabricated by Pb ion implantation and subsequent high-temperature annealing (2100℃) under high pressure (7.7 GPa). Their optical properties were characterized by photoluminescence at varying temperatures down to 5.7 K. We observed intense emission peaks at 550 and 554 nm with a large splitting of approximately 3900 GHz. The two lines are thought to correspond to the zero phonon line (ZPL) of PbV centers with split ground and excited states. A cubic trend of the ZPL width was observed while varying temperature. We performed polarization measurements of the two lines in a single PbV center, showing nearly orthogonal dipole polarizations. These optical measurements strongly indicate that the PbV center possesses $D_{3d}$ symmetry in the diamond lattice. The observed large ground state splitting significantly suppresses the phonon-mediated transition, which causes decoherence of the electron spin state of the group IV color centers in diamond, expecting a long spin coherence time at a temperature of ~8 K.

The rapid development of quantum information brings great demand for the transfer of quantum states among separated quantum nodes[1,2]. Color centers in diamonds are expected to be suitable candidates for quantum nodes[3–7]. Nitrogen-vacancy (NV) centers in diamond have been intensively studied because of their excellent spin coherence time[8–12]. However, the NV center has a low concentration on the zero phonon line (ZPL) against the total fluorescence, at ~4%, and an unstable optical transition against external noise[13]. To overcome these issues, group IV element-based color centers in diamond, such as silicon-vacancy centers (SiV)[13–15], germanium-vacancy centers (GeV)[16–18], and tin-vacancy centers (SnV)[19–22], have been attracting interest because they possess large ZPLs and high resistance to external noise due to their structural symmetry[23,24]. Nevertheless, their spin coherence times are limited by phonon-mediated transitions within the ground states[25–28]. Sub-Kelvin cooling of an SiV center has been reported to achieve a long spin coherence time of over 10 ms[29]. As another approach, a heavier group IV element of tin was used to fabricate SnV centers to suppress the phonon-mediated transition with a large ground state splitting[19]. Indeed, superior spin properties have been demonstrated for an SnV center[22] at a similar temperature compared with SiV[27] and GeV[28] centers. Then, the utilization of a heavier atom of lead (Pb) is a straightforward strategy to obtain superior spin properties at an elevated temperature. To date, a few studies have been reported for the fabrication of lead-vacancy (PbV) centers in diamond. However, the optical transition energy of the PbV center has not been determined consistently: values have been determined at ~550 nm[30,31], or ~520 nm[32].

Polarization of the fine structure is a powerful method to relate the optical properties to the atomic symmetry of the color center. The group IV color centers possess a split-vacancy configuration, with an impurity atom at an interstitial site between two carbon vacancies. This structure corresponds to $D_{3d}$ symmetry with inversion symmetry, producing a four-level energetic structure composed of split ground states and split excited states[33], as shown in Fig. 1. Upon optical measurements on a (001) diamond crystal, a pair of two spectral lines, referred to as C- and D-peaks, show orthogonal polarization characteristics[34,35]. Theoretically, the PbV center is also predicted to have $D_{3d}$ symmetry[32,33]. Thus, the observation of the polarization provides important insight to determine the characteristics of the PbV center. High-quality color centers are essential to precisely measure the optical properties. However, ion implantation of the heavy atom of Pb leads to the formation of numerous defects, leading to large strain around the color centers inside the diamond crystal, which are not efficiently recovered by usual annealing at ~1200°C, as shown for SnV[36]. For high-quality fabrication, in this study, we prepared PbV centers by performing high-pressure and high-temperature (HPHT) treatment at 2100°C, which was demonstrated for the fabrication of SnV quantum emitters[19]. We observed temperature-dependent photoluminescence (PL) in the ensemble state and fine structure polarization in a single state. Furthermore, the excited state lifetime and saturation intensity were estimated for a single PbV center.

Experimental conditions
We fabricated PbV centers in diamond by Pb ion implantation into IIa-type (001) single-crystal diamond substrates. Ion implantation was performed with doses of $1 \times 10^{13}\ cm^{-2}$ (high-dose sample) and $2 \times 10^{9}\ cm^{-2}$ (low-dose sample) at acceleration energies of 150 or 330 keV, leading to projected depths of 34 nm or 57 nm, respectively, calculated by SRIM[37].

Subsequently, the samples were annealed at up to 2100 °C under a high pressure of 7.7 GPa[19]. High temperature was used to reduce the inhomogeneous distribution by effectively recovering lattice damage induced by ion implantation[19,36]. Graphite becomes the stable phase for carbon under low pressure; thus, high pressure was applied to stabilize the diamond phase at high temperature. The surface morphology of the samples is shown in Supplemental Material. We performed optical measurements utilizing a home-built confocal fluorescence microscope equipped with a temperature-controlled cryostat (Supplemental Material). The confocal fluorescence maps were collected with the Qudi Python suite[38].

Results and Discussion

**PL spectra.** The PL spectra of both the high-dose and low-dose samples were acquired at room temperature and a low temperature of 5.7 K under 515 nm excitation, as shown in Fig. 2. At room temperature, four prominent peaks are observed at 552 ($C$), 556 ($D$), 591 ($\alpha$), and 717 nm ($\gamma$) for the high-dose sample, labeled in order of wavelength. When cooling down to 5.7 K, one more peak is visible at ~593 nm ($\beta$). The doublet at approximately 550 nm agrees with previous studies[30,31], which are thought to originate from the PbV center. Distinguishable splitting of the C- and D-peaks can be observed even at room temperature despite their thermal broadening. In contrast, another doublet at approximately 591 nm shows a different behavior while varying the temperature: the $\beta$ peak becomes invisible at approximately 150 K before it merges into the $\alpha$ peak (Supplemental Material). Meanwhile, the spectrum of the low-dose sample gives a more reliable spectrum of the PbV center. There was only one prominent doublet observed at approximately 550 nm regardless of temperature, in agreement with the high-dose sample. A sharp line between them is the first-order Raman scattering of diamonds under 515 nm excitation[39]. The fluorescences at 575 nm and 637 nm come from neutrally and negatively charged NV centers in diamond[40], respectively. Note that we also observe a broad emission at around 590 nm, corresponding to the $\alpha$ and $\beta$ peaks, which are considered to be phonon sideband of PbV center according to calculation[33]. Whereas, the $\gamma$ peak was not prominently observed in the low-dose sample, which might be related to another defect formed by heavy ion implantation.

Considering the four-level transition defined by the $D_{3d}$ symmetry of the group IV vacancy center in diamond[19,34], as shown in Fig. 1, we consider the spectral doublet at approximately 550 nm (C- and D-peaks) as the ZPL of the PbV centers in diamond, which originates from the transition from the lower excited state $|e_-^u\rangle$ to two ground states $|e_+^g\rangle$ and $|e_-^g\rangle$ (Fig. 1(b)). The separation of the two peaks corresponding to the ground state splitting is estimated to be 3952 GHz for the high-dose sample and 3878 GHz for the low-dose sample at 5.7 K, which agrees with the theoretical calculation of 4385 GHz[33]. These values are much larger than those of other group IV emitters since the heavy Pb atom leads to a large spin-orbit interaction that dominantly determines the ground state splitting.

We adopted another excitation laser of 355 nm to investigate the wavelength range of approximately 520 nm where Pb-related peaks have been reported before[32], but we did not observe any obvious emission (Supplemental Material). This might be caused by the different excitation laser wavelength, which modifies the charge state of color centers, and/or slightly different structures depending on the fabrication process. Therefore, hereinafter, we focus on optical investigations on the doublet at approximately 550 nm.

**Temperature dependence of optical transitions.** We conducted temperature-dependent PL measurements on ensemble PbV centers in the high-dose sample from 5.7 to 263 K, as shown in Fig. 3(a). The C- and D-peaks are seen at all temperatures. As the temperature increases, the C- and D-peaks show a clear redshift and broadening. Figure 3(b,c) depicts the temperature dependence of the full width at half maximum (FWHM) of the two peaks. Note that the peak widths were above the spectrometer limit (~60 GHz) due to inhomogeneous broadening in the high-dose ensemble centers. A $T^3$ curve is well fitted with both the C- and D-peaks. The cubic trend has been also reported in the other group-IV color centers[18,36,41]. This temperature dependence could be attributed to a joint result of inhomogeneous distribution and homogeneous expansion. Inhomogeneous expansion is inevitable during investigation on ensemble centers[42,43] in which each emitter possesses slightly different wavelengths. This phenomenon was observed even for two PbV centers in the low-dose sample (Supplementary Material). The latter originates from the second-order electron-phonon transition owing to the Jahn–Teller effect[41,44]. The shift of both peaks is very well described with a combination of $T^2$ and $T^4$ power laws, as shown in Figs. 3(d,e). This temperature dependency also agrees well with the previous report on the SnV centers[36]. Notably, two other peaks, referred to as A- and B-peaks, were not observed in this study (Fig. 3a). The excited state splitting increases as the atomic number of the group IV impurity increases owing to the spin-orbit interaction: 6920 GHz is predicted for the PbV center[33], much higher than those of other group IV emitters. Therefore, high temperature is required to excite the electron to the upper branch in the excited state for the PbV center and then to observe the A- and B-transitions, while high temperature accelerates peak broadening, reducing the peak intensity in return, which is thought to inhibit the observation of the two peaks.

**Single PbV center.** We found a single PbV center in the low-dose sample and investigated the optical properties. Figure 4(a) shows confocal florescence microscopy imaging of an isolated PbV center at 5.7 K. The measurement was conducted with a bandpass filter (BPF) with a 23 nm band width at approximately 554 nm. Notably, the background signal around the emitter is mainly caused by first-order Raman scattering. Both the C- and D-peaks are again observed from the bright spot (Fig. 4(b)). Then, we performed the Hanbury Brown and Twiss measurement[45] to confirm the single photon emission from the observed spot. Figure 4(c) shows the second-order autocorrelation function, $g^2(\tau)$, at an excitation laser power of 0.2 mW. The optical center exhibits an antibunching dip at $\tau = 0$ ns. Even before subtracting the background emission, $g^2(0)$ is below 0.5, which provides proof of single photon emission[46] from the PbV center. Notably, the background-corrected $g^2_{corr}(0)$ reaches almost zero (Supplemental Material). We fit the curve in Fig. 4(c) to the function[21] of $g^2(\tau) = 1 - c[(1+b)e^{-|\tau|/\tau_1} - be^{-|\tau|/\tau_2}]$, where $b, c, \tau_1, \tau_2$ are fitting parameters. $\tau_1$ yields an estimation of the excited state lifetime of a florescent structure at a low excitation power[47]. The excited state lifetime of the observed single PbV center was estimated to be ~3.7 ns, which is consistent with another experimental result[31]. The fluorescence intensity is measured as a function of the excitation laser power, as shown in Fig. 4(d). The power dependence is plotted after subtracting the background emission (data before subtraction are shown in Supplementary Material) and then fitted with an equation[48] of $I = I_\infty P/(P + P_{sat})$, where $I_\infty$ and $P_{sat}$ represent the saturation intensity (~222 kcps) and saturation power (~1.8 mW), respectively.

**Polarization of single PbV center.** The polarization of the C- and D-peaks of a single PbV center was investigated at a temperature of 5.7 K. PL spectra of the fine structure were recorded by a spectrometer while rotating a half wave plate placed in the detection path. Figures 5(a) and (b) show the polarization of the C- and D-peaks, respectively. The polarization of the dipole is determined by the fine structure of the color center[34,35]. For the C-peak from the $|e_-^u\rangle \to |e_-^g\rangle$ transition (Fig. 1(b)), linearly polarized light of the Z dipole along the main axis is emitted, while the D-peak from the $|e_-^u\rangle \to |e_+^g\rangle$ transition has an XY-plane polarization perpendicular to the high symmetry axis. Investigated through a (001) surface, the projections of both C- and D-peaks are expected to be orthogonal to each other[34,35], which matches the observed polarization in Fig. 5. This orthogonal polarization serves as a strong indication for the $D_{3d}$ symmetry of the PbV center. To further characterize the dipole of the single PbV center in diamond, we estimated the visibility of both peaks, which is defined as[49] $V = (I_{max} - I_{min})/(I_{max} + I_{min})$, where $I_{max}$ and $I_{min}$ represent the maximum and minimum intensities. The visibility of the C-peak is derived as 94%, while that of the D-peak is estimated to be 63%. The observed visibility of the C-peak agrees well with the ideal linear Z dipole (100%) (Supplemental Material).

**Expected spin property.** The spin coherence time of group IV color centers is dominantly limited by the photon-mediated transition in the ground state, which depends on the measurement temperature (T) and ground state spitting ($\Delta_{gs}$). Here, the expected spin property of the PbV center is discussed according to the phonon-mediated transition rate[19,41] given as $\gamma_+ = 2\pi\chi\rho\Delta_{gs}^3[\exp(h\Delta_{gs}/kT) - 1]^{-1}$, where $\chi$ and $\rho$ are proportionality constants. h and k are the Planck and Boltzmann constants, respectively. Figure 6 shows the calculated phonon-mediated transition normalized with the value for a strained SiV center (SiV-II) at a temperature of 0.4 K[29], where a long spin coherence time was obtained. While the SnV center is expected to have a comparable spin property at ~2 K[19], the even larger splitting in the ground state of the PbV center further increases the temperature to ~8 K. The current simulation strongly suggests that the PbV center will be an important system to construct a quantum light-matter interface, which is the key element for the application of quantum networks.

Conclusion
We demonstrated the formation of the ensemble and single PbV centers in diamond via ion implantation and HPHT annealing. A clearly separated florescent doublet was observed at 550 nm and 554 nm, leading to large ground state splitting of~3900 $GHz$. The polarization of a single PbV center leads to a strong indication of the $D_{3d}$ symmetry. The utilization of the heavy element increased the ground state splitting owing to the larger spin-orbit interaction compared with other group IV centers. Thus, a long spin coherence time is expected at ~8 K for the PbV center.

During the preparation of the manuscript, we became aware of another study presenting PL spectra of PbV centers in diamond from 4 to 300 K at various excitation lasers[50]. Their PbV centers were formed at a lower annealing temperature of 1200℃.

Acknowledgment

We thank Yasuyuki Narita for the SRIM calculation and Shinji Nagamachi for supporting in ion implantation. This work is supported by the Toray Science Foundation and the MEXT Quantum Leap Flagship Program (MEXT Q-LEAP) Grant No. JPMXS0118067395.

Author information
Corresponding Author
E-mail: iwasaki.t.aj@m.titech.ac.jp
ORCID: orcid.org/0000-0001-6319-7718

Notes:
The authors declare no competing financial interests.

Supporting Information:
The followings are provided in the supplemental material.
1. Optical measurement system, sample preparation, and surface morphology
2. Additional PL spectra
3. Background corrected autocorrelation function and saturation curve of a single PbV center
4. Estimation of polarization visibility
5. Observation of strained double PbV centers

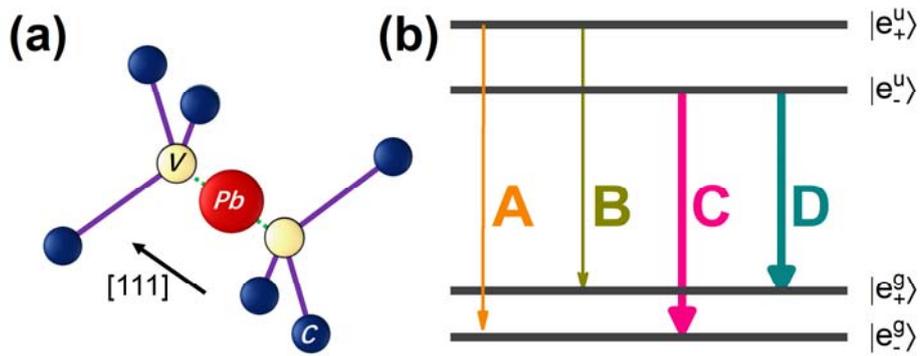

Figure 1. PbV center in diamond. (a) Expected atomic structure of the PbV center in the diamond lattice. The main axis is along the [111] direction. The observations in this study were performed along the [001] direction. (b) Fine structure with four possible emissions, labeled A-D. Although each level is a spin doublet[41], as an example, here only the spin up levels are illustrated.

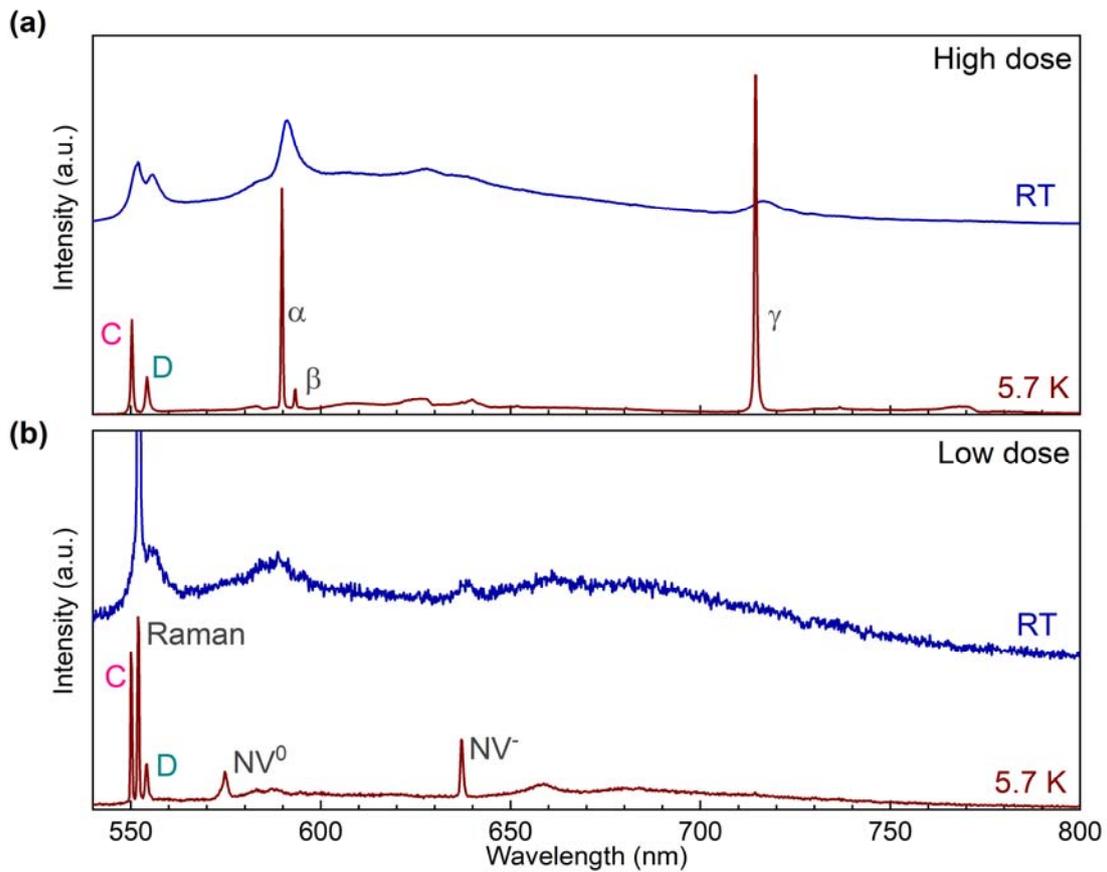

Figure 2. PL spectra under 515 nm laser excitation at 5.7 K (red) and at room temperature (~300 K) (blue). (a) High-dose sample. There were five prominent peaks observed at 5.7 K: 550 nm (C), 554 nm (D), 590 nm ($\alpha$), 593 nm ($\beta$), and 717 nm ($\gamma$). (b) Low-dose sample. There were two prominent peaks at 550 nm (C) and 554 nm (D) observed at 5.7 K (apart from the Raman scattering peak and NV peaks). The number of the emitters at the measurement spot was not clarified.

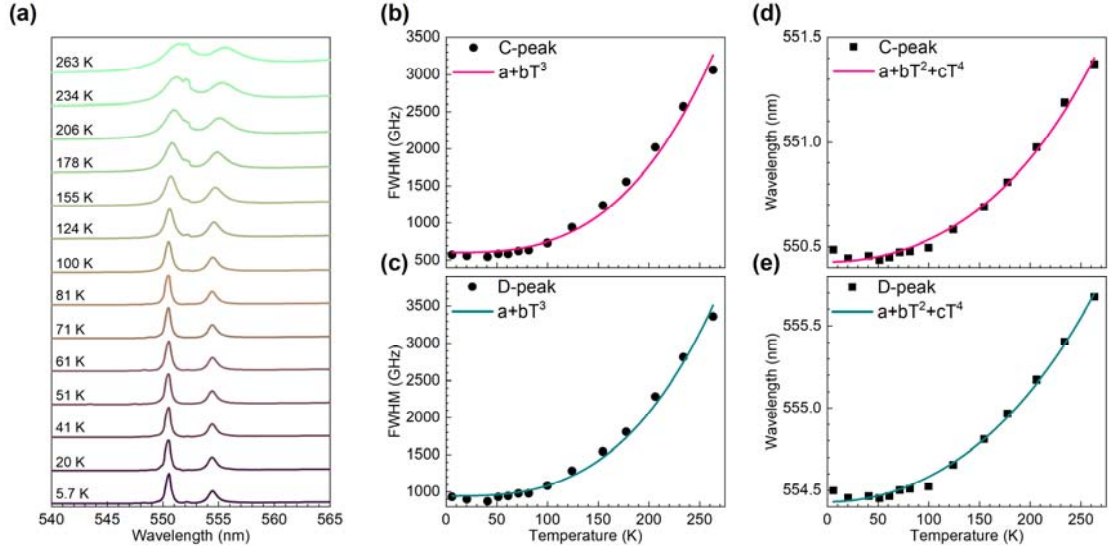

Figure 3. Temperature dependence of PL spectra under 515 nm laser excitation from the high-dose sample. (a) Fine structure of PbV centers from 5.7 to 263 K. A peak between the C- and D-peaks corresponds to the first Raman scattering from diamond. Temperature dependence of the linewidth of the (b) C-peak and (c) D-peak. The solid lines are fitted with $T^3$ dependence. The peak wavelengths of the (d) C-peak and (e) D-peak fitted with $T^2 + T^4$.

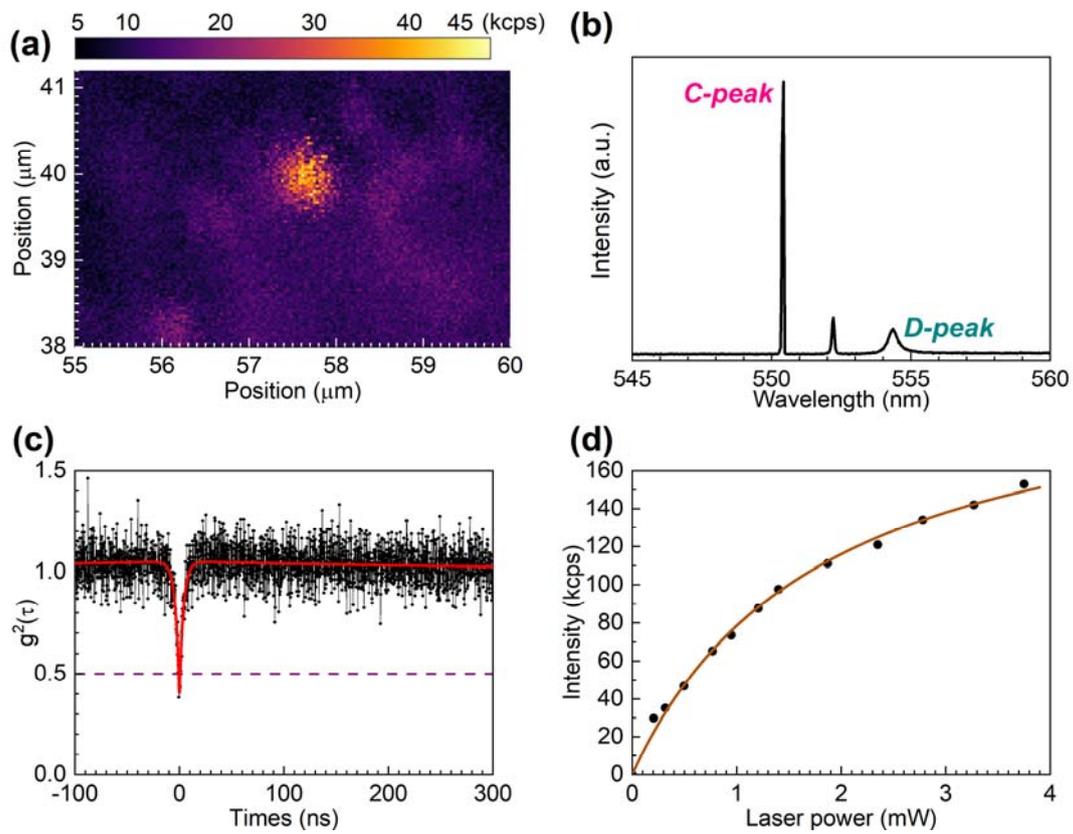

Figure 4. Single PbV center in diamond. (a) Plane-view confocal fluorescence microscope image obtained using a BPF (23 nm width around the ZPL). Background emission comes from Raman scattering of diamond. (b) PL spectrum of single PbV center. (c) Second-order autocorrelation function $g^2(\tau)$ without background correction at a laser power of 0.2 mW. (d) Saturation curve of ZPL emission with subtracting background intensities. All measurements were performed at 5.7 K.

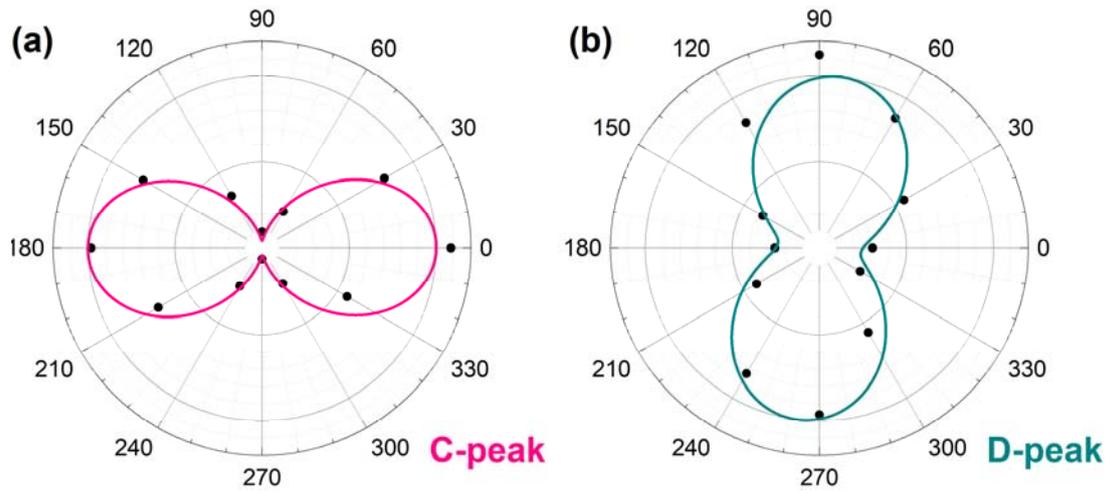
Figure 5. Polarization of the ZPL fine structure of a single PbV center. (a) C-peak and (b) D-peak. The radius represents the florescent intensity in the spectrum.

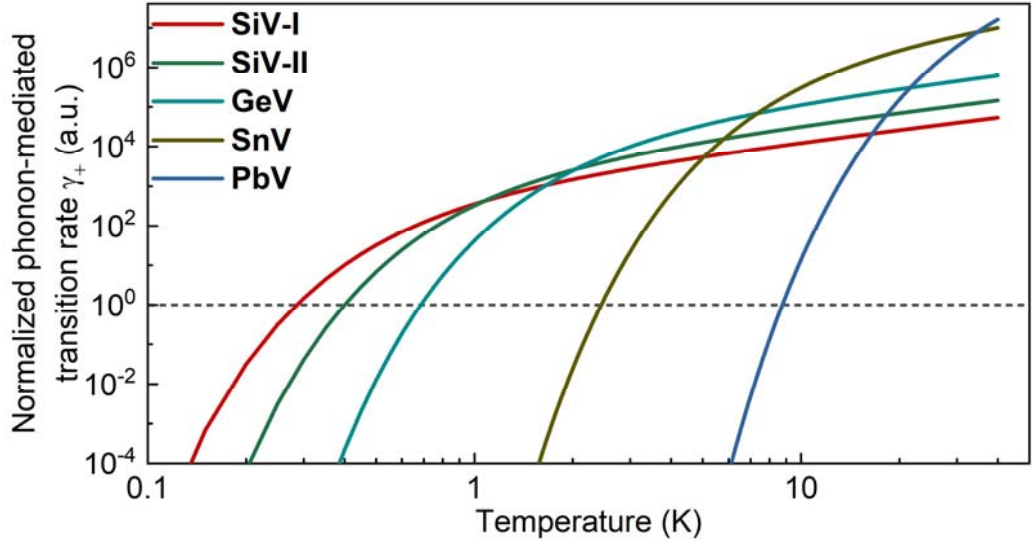

Figure 6. Phonon-mediated transition rate $\gamma^+$ of SiV-I (48 GHz)[14], SiV-II (80 GHz)[29], GeV (169 GHz)[17], SnV (850 GHz)[19] and PbV (3878 GHz). The parentheses represent the corresponding ground state splitting. The transition rate $\gamma^+$ is normalized to the value for SiV-II at a temperature of 0.4 K, mentioned as a dashed line.

# Supplemental Materials for

# Low-Temperature Spectroscopic Investigation of Lead-Vacancy Centers in Diamond Fabricated by High-Pressure and High-Temperature Treatment

Peng Wang[1], Takashi Taniguchi[2,] Yoshiyuki Miyamoto[3], Mutsuko Hatano[1], and Takayuki Iwasaki[1, *]

[1]Department of Electrical and Electronic Engineering, School of Engineering, Tokyo Institute of Technology, Tokyo 152-8552, JAPAN
[2]Research Center for Functional Materials, National Institute for Materials Science, Tsukuba 305-0047, JAPAN
[3]Research Center for Computational Design of Advanced Functional Materials, National Institute of Advanced Industrial Science and Technology, Tsukuba 305-8568, JAPAN

### 1. Optical measurement system, sample preparation, and surface morphology

Optical measurements were performed by using a home-built confocal fluorescence microscope (CFM) system equipped with a cryostat and a 515 nm green laser. The florescence from the PbV centers was filtered with a 554 nm bandpass filter and then split by a nonpolarizing beam splitter to be collected with two avalanche photodiodes (APDs). The PL measurements were carried out by a spectrometer with a resolution of approximately 60 GHz for 1800 g/mm grating. The polarization was obtained by the combination of a half-wave plate and a polarizer. The samples used in this study were fabricated by annealing at 2100℃ and 7.7 GPa after Pb ion implantation. High pressure was applied in a pressure transmitting medium (cesium chloride) in a belt-type high-pressure system[1].

It was found that the surface morphology of the diamond substrates was affected by the HPHT annealing process. Figure S1 shows the surface morphology of the high-dose sample observed by optical microscopy (OM), atomic force microscopy (AFM), and CFM. In the OM image, we see many black dots, which correspond to pyramid-like structures with a height of 0.4~2.2 $\mu m$ in AFM (Fig. S1b). Emission from the PbV centers was observed over the whole scanned region (Fig. S1c), meaning that the implanted Pb atoms were not removed from the surface but remained during the HPHT annealing process. This fact indicates that pyramid-shaped structures are newly formed on the surface and are not formed by etching. On the other hand, previously, the surface of a diamond sample with Sn ion implantation suffered from etching, removing the Sn atoms to a large extent[2]. As shown in Fig. S2(a,c), etching was also likely to occur in the low-dose sample, leading to a roughness of 28 nm, while many pyramid-like structures were observed on another half of the same substrate (Fig. S2a,b). The single PbV center investigated in the main text was found at the region near the center of the sample. It is thought that the epitaxial growth of the structures and surface etching would simultaneously occur during the HPHT anneal, and the temperature gradient in the high-pressure cell and the amount of impurities such as carbon from the graphite heater and/or diamond substrates for growth and water for etching determine which process becomes dominant on the surface.

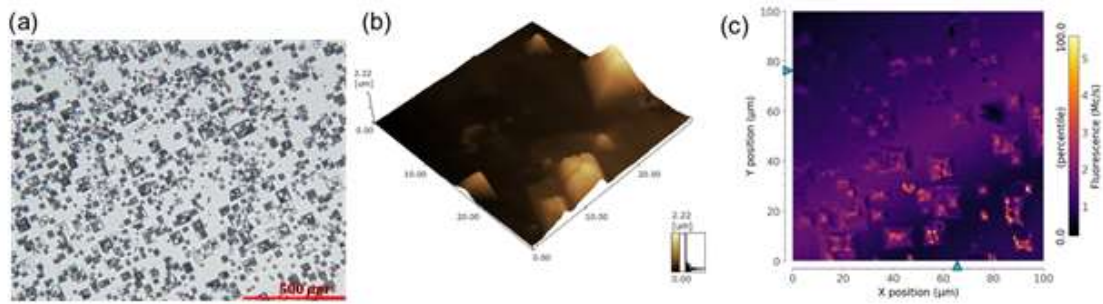

Fig. S1 Surface morphology of the high dose sample. (a) OM image. (b) AFM image. (c) CFM image for the PbV emission. Inhomogeneous intensity was caused by the slightly inclined sample surface attached on the sample holder.

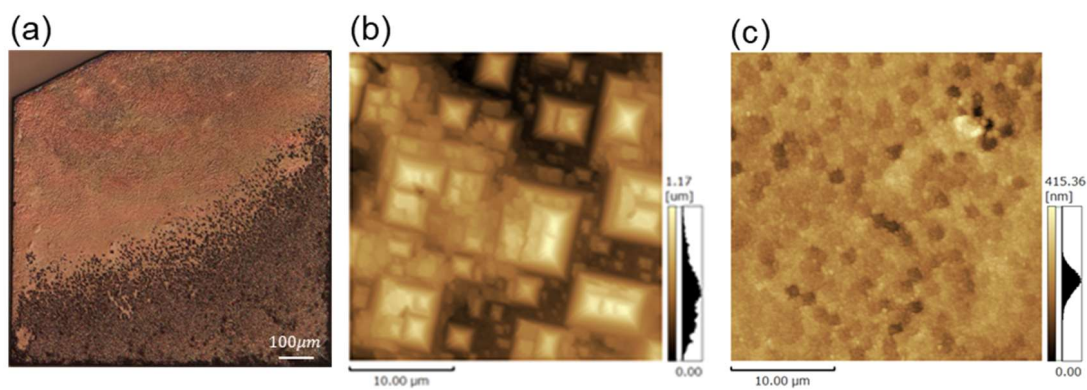

Fig. S2 Surface morphology of the low dose sample. (a) OM image. (b) AFM image of a pyramid area (bottom right in panel (a)). (c) AFM image of a flatter region (upper left in panel (a)).

## 2. Additional PL spectra

We attributed a peak at ~552 nm to the diamond Raman peak under 515 nm excitation. We confirmed this by comparing spectra from the PbV center and background emission (Fig. S3).

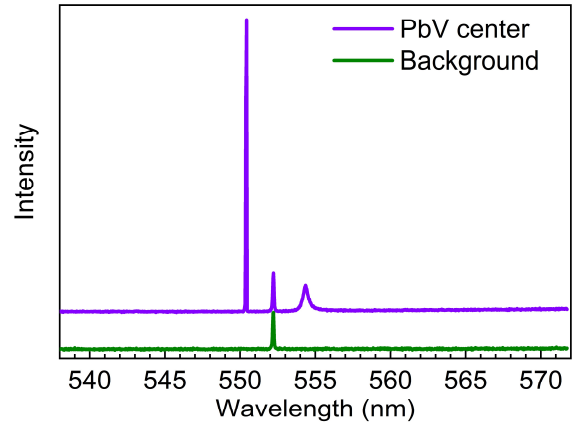

Fig. S3 PL spectra from PbV center (purple) and background emission (green).

Figure S4 shows the temperature dependence of the $\alpha$ and $\beta$ peaks of the high-dose sample. Compared to the normalized $\alpha$ peak at 590 nm, the $\beta$ peak at approximately 593 nm decreased dramatically with increasing temperature and became invisible at ~150 K. The peaks at this wavelength range might be phonon sideband of PbV according to the first-principles calculation[3].

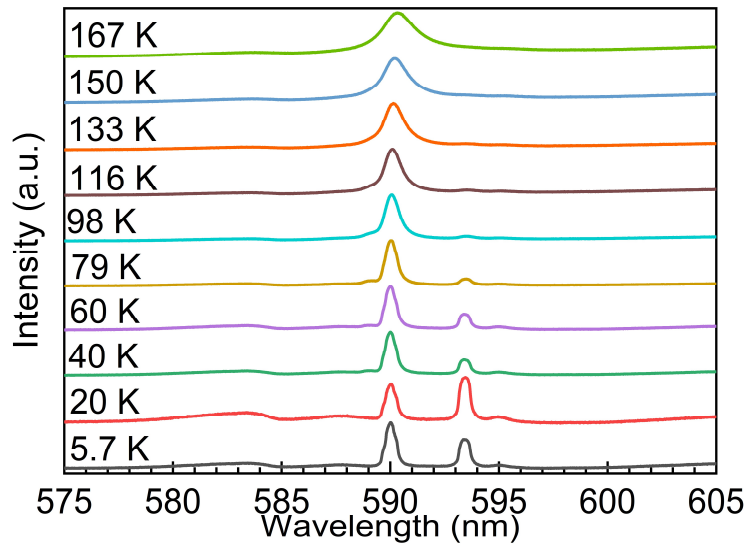

Fig. S4 Temperature dependence of the $\alpha$ and $\beta$ peaks.

To examine the presence of a peak at 520 nm observed in a previous study[4], we performed room temperature PL measurement using a micro-Raman system equipped with an excitation laser of 355 nm (Fig. S5). No prominent peak was observed in this range, while again, two peaks corresponding to the C- and D-peaks at 552 nm and 556 nm were observed.

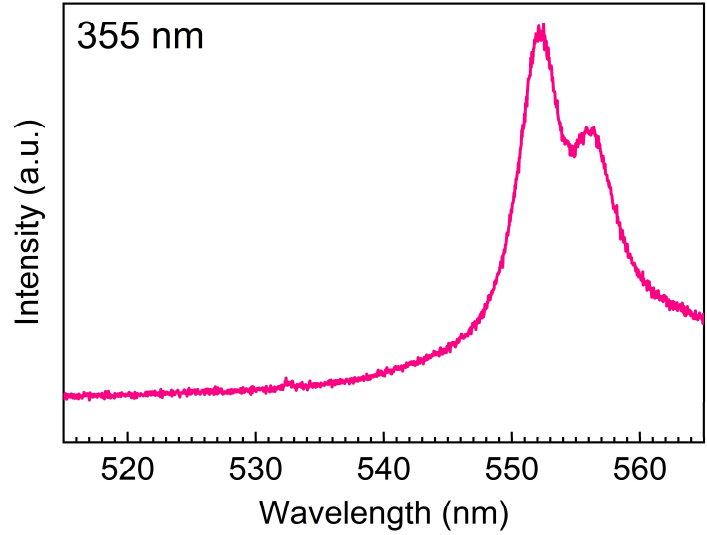

Fig. S5 PL spectrum of the high-dose sample under 355 nm excitation (room temperature).

**3. Background corrected autocorrelation function and saturation curve of a single PbV center**

We exhibited the second-order autocorrelation function, g²(τ), of a single PbV center in the main text, showing the antibunching $g^2(0) < 0.5$ even before background correction. Here, we corrected the background emission by using a relation[5] $g^2_{corr}(\tau) = [g^2(\tau) - (1-\rho^2)]/\rho^2$. $\rho$ is defined as $\rho = S/(S+B)$, where S and B denote the fluorescence intensity of the PbV center and background intensity, respectively. For the single PbV center, we found $\rho = S/(S+B) \approx 0.76$. As shown in Fig. S6, $g^2_{corr}(0)$ goes to almost zero, which again indicates the single photon emission. The observation of slight bunching indicates the existence of a shelving state in addition to the ground and excited states[6].

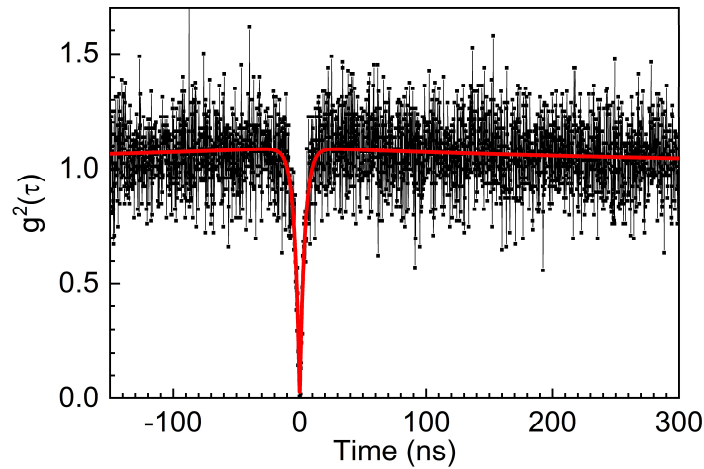

Fig. S6 Background corrected second-order autocorrelation of a single PbV center.

Figure S7 shows the saturation curve of a single PbV center, including the total, background, and background subtracted signal intensities. The blue line represents the total fluorescence intensity from the single PbV center in the diamond sample, while the pink line represents the background emission obtained from the dark region where no spectral doublets were observed. Both were collected with 554 nm bandpass filter. Then, their difference leads to pure florescence from the PbV center, plotted in brown, which is shown in the main text.

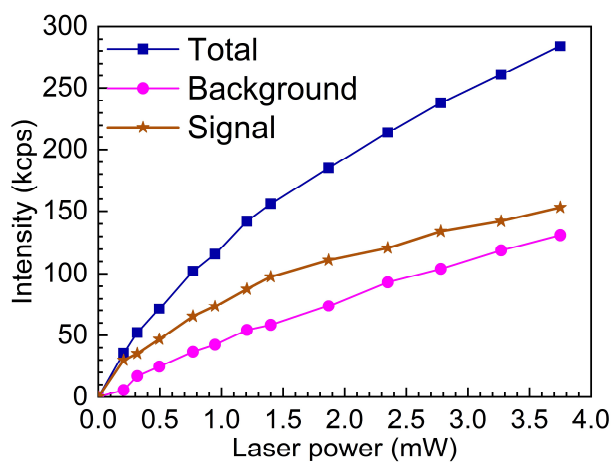

Fig. S7 Saturation curve of a single PbV center.

## 4. Estimation of polarization visibility

The visibility of the polarization of the $D_{3d}$ symmetry color center in diamond is estimated by considering the projection of the dipole onto the (001) surface[7], as shown in Fig. S8. On the (001) surface, both $u$ and $v$ denote the orthogonal <110> directions. The yellow object represents the main Z axis of the group-IV color center oriented in the ⟨111⟩ direction. The XY plane is perpendicular to the Z axis. The projections of the X, Y, and Z dipoles onto the (001) surface are given as

$$u_X = 0.577, \quad v_X = 0$$

$$u_Y = 0, \quad v_Y = 1$$

$$u_Z = 0.816, \quad v_Z = 0$$

Thus, for a purely linear dipole in the Z-axis of the C-transition, perfect linear polarization with 100% visibility is expected. Our observation of 94% agrees well with this calculation. When assuming a combination of linear X and Y dipoles with the same intensity for the D-transition, the visibility is estimated to be 26.8%. The obtained visibility of the D-peak (63%) is higher than the estimation, which implies that the Y dipole might become stronger[7].

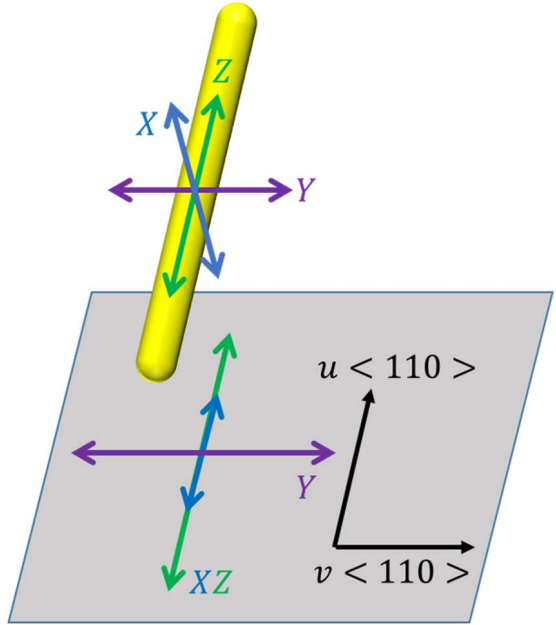

Fig. S8 Schematic of projection of the dipole onto the (001) surface.

## 5. Observation of strained double PbV centers

In the low-dose sample, we observed double PbV centers. The background corrected second-order autocorrelation function at 0 ns is close to 0.5 (Fig. S9a). $g^2(0)$ gives the information on the number of emitters (n) with the relation[8] of $g^2(0) = 1 - 1/n$, indicating that the two PbV centers exist at the measured position. In a low temperature PL spectrum at 5.7 K, we see two sets of C- and D-peaks, whose splitting was 3876 and 3898 GHz, respectively (Fig. S9b). Again, the middle peak at 552 nm was Raman scattering under 515 nm excitation. The split of these peaks is expected to be attributed to the strain environment around the PbV centers. This observation will lead to the tuning of the fluorescence wavelength through strain engineering, as demonstrated for SiV[9] and GeV centers[10].

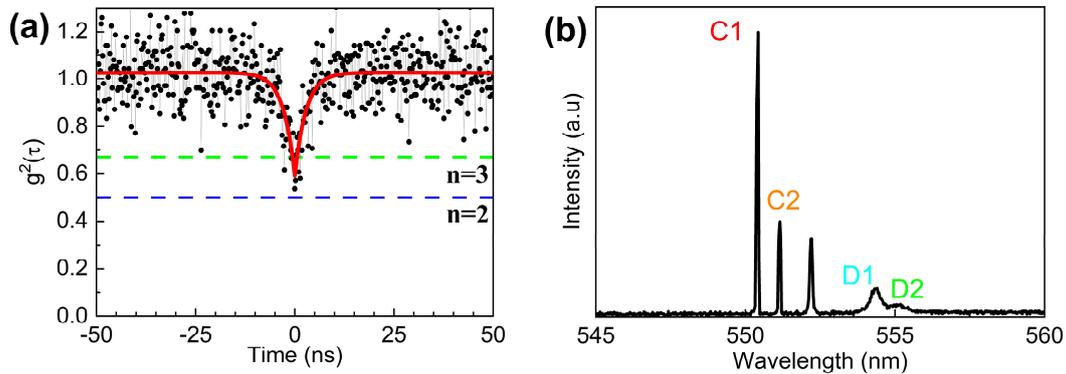

Fig. S9 Observation of double PbV centers. (a) Background corrected second order autocorrelation. (b) PL spectrum of double PbV centers at 5.7 K.